\documentclass[twocolumn,showpacs,preprintnumbers,amsmath,amssymb,groupedaddress,prb]{revtex4}

\usepackage{graphicx}
\usepackage{dcolumn}
\usepackage{bm}
\usepackage{epsfig}
\usepackage{subfigure}
\def\e{\begin{equation}}
\def\f{\end{equation}}
\def\=#1{\overline{\overline #1}}

\def\-#1{{\bf #1}}
\def\.{\cdot}

\def\.{\cdot}
\def\x{\times}

\begin{document}

\title{Local constitutive parameters of metamaterials}

\author{Constantin R. Simovski}
\affiliation{
Physics
Department, State University of Information Technologies, Mechanics
and Optics, Sablinskaya 14, 197101, St. Petersburg, Russia}

\begin{abstract}
The reason of the non-locality of constitutive (material) parameters
extracted in a usual way from the reflection-transmission
coefficients of composite slab at moderately low frequencies is
explained. The physical meaning of these parameters is clarified.
Local constitutive parameters of metamaterial lattices are discussed
and their existence at moderate frequencies is demonstrated. It is
shown how to extract local material parameters from the dispersion
characteristics of an infinite lattice and from reflection and
transmission coefficients of metamaterial layers.
\end{abstract}


\maketitle
\section{Introduction}
In the modern scientific  literature the so-called {\it
metamaterials} \cite{Caloz,Ramakrishna} correspond to a lot of
exciting results and a lot of mess. Metamaterials (MTM) are often
defined as media which possess constitutive (material) parameters
not observed in nature. Thanks to these material properties, many
exotic phenomena were predicted and discovered in the MTM structures
(a detailed overview is presented in \cite{Caloz}). As a rule, MTM
are presented by lattices of reciprocal optically small resonant
particles, such as complex-shape metal inclusions (in different
frequency ranges), plasmonic and polaritonic nano-spheres and
nano-wires (at infrared waves and in the visible). Particles
characteristic size $\delta$ and maximal period of lattices $a$ at
the frequency of particle resonance are usually assumed to be much
smaller than the wavelength $\lambda$ in the lattice matrix. This
assumption is usually considered as allowing one to homogenize MTM
and to explain their exotic properties through material parameters
(MP). However, it is important that $\delta$ and $a$ in MTM though
small are not negligible with respect to $\lambda$. Practically,
they lie within the frequency band \e 0.01<{(a,\delta)\over\lambda}<
0.1.\label{eq:fb}\f As a result, the effective wavelength in MTM (it
can be expressed though the lattice eigenwave wavenumber $q$ as
$\lambda_{\rm eff}=2\pi/q$) being shortened at the particle
resonance compared to $\lambda$ can approach to $a/2$. At such
frequencies the lattice spatial resonance (often called as
\emph{Bragg's resonance}) holds. Then the Bragg (staggered) mode
which has complex wavenumber $q=\pi/a+j{\rm Im}(q)$ can be excited
in the lattice. Within the band of the staggered mode the
homogenization of the metamaterial lattice is meaningless. If one
formally introduces material parameters for such a regime they
cannot satisfy to the physical conditions of locality. Basic
physical limitations which are equivalent to the concept of locality
are as follows (see e.g. in \cite{Landau}):
\begin{itemize}
\item
passivity (for the temporal dependence $e^{-i\omega t}$ it implies
${\rm Im}(\varepsilon)>0$ and ${\rm Im}(\mu)>0$ simultaneously at
all frequencies, for $e^{j\omega t}$ the sign of both ${\rm
Im}(\varepsilon)$ and ${\rm Im}(\mu)$ should be negative);
\item
causality (in the region of negligible losses the frequency
derivatives of both ${\rm Re}(\varepsilon)$ and ${\rm Re}(\mu)$
should be positive);
\item
absence of radiation losses (it was postulated for periodic infinite
structures in \cite{Man}, then derived in \cite{Man1}, now this
principle is often attributed to work \cite{Sipe}).
\end{itemize}
Constitutive parameters of MTM are non-local at certain frequencies
inside the resonant band of the particle. From it one often deduces
that it is impossible to introduce local MP over the whole resonant
band. In the present paper we will show that this opinion is wrong.

It is evident that the staggered mode of a lattice does not cover
the whole region of moderately low frequencies \ref{eq:fb}, and
there are bands where $a< \lambda_{\rm eff}/2$. There we should
expect the locality of measured or simulated MP if they are measured
or simulated properly). However, inspecting well-known works
\cite{Smith,OBrien0,Soukoulis11,OBrine2,Smith1,Huang,Soukoulis2,Yen}
(and many others) devoted to the extraction of material parameters
(MP) from measured or simulated reflection ($R$) and transmission
($T$) coefficients of a metamaterial slab one observes another
result. At least one of two extracted MP in all these works violates
all locality conditions over the whole frequency range \ref{eq:fb}.
It will be shown below that the reason of it is the different
physical meaning of these MP than the meaning of local constitutive
parameters introduced in the quasistatic theory of lattices (e.g.
\cite{Born}).  In \cite{Pendrynew} it was properly noticed that MP
introduced for orthorhombic lattices in \cite{Pen,Pen1} are
non-local \emph{by definition}. In fact, these MP are exactly the
same as MP measured or simulated in
\cite{Smith,OBrien0,Soukoulis11,OBrine2,Smith1,Huang,Soukoulis2,Yen}.
In the present paper these non-local MP will be complemented by
local ones extracted from same $R$ and $T$ for the same frequency
range defined by \ref{eq:fb}.

In \cite{Smith} one proposed to extract MP of lattices of thickness
$d$ as if these were filled by a continuous medium with parameters
$\varepsilon_{\rm eff},\ \mu_{\rm eff}$. Then $R$ and $T$ for the
normal incidence take form: \e R={r(1-e^{-2jq_{\rm eff} d})\over
1-r^2e^{-2jq_{\rm eff} d}},\quad T={e^{-jq_{\rm eff} d}(1-r^2)\over
1-r^2e^{-2jq_{\rm eff} d}}. \label{eq:rt}\f Here $q_{\rm
eff}=\omega\sqrt{\varepsilon_0\mu_0\varepsilon_{\rm eff}\mu_{\rm
eff}}\equiv k_0\sqrt{\varepsilon_{\rm eff}\mu_{\rm eff}}$ is the
wavenumber of the medium filling the layer. Also in these formulas
$r$ is the reflection coefficient from semi-infinite medium
expressed through $Z$ (medium wave impedance normalized to that of
free space) as \e r={Z-1\over Z+1},\quad Z={\sqrt{ \mu_{\rm
eff}\over\varepsilon_{\rm eff}}}. \label{eq:standard}\f

\section{Local and non-local material paramerets}

In works \cite{SimPRB,Sim,EuPJ,EuPJ1} it was shown that the
approximation of a uniform slab filled by effective continuous
medium is too rough within the frequency range \ref{eq:fb}. This is
so even for non-resonant inclusions studied in
\cite{SimPRB,Sim,EuPJ,EuPJ1}, and, moreover, it is so for resonant
ones. However, this observation does not mean that the
homogenization of slabs at moderately low frequencies is impossible
at all. In these works two slightly different algorithms of
homogenization were developed. The original composite slab of
thickness $d$ was replaced by a three-layer structure. These
effective layers are assumed to be filled by different continuous
media. The central layer
was described through local constitutive parameters $\varepsilon_L$
and $\mu_L$ which were calculated through the known polarizabilities
of particles. Two \emph{transition} layers of small thickness
$d_t\approx \delta$ were described by special anisotropic material
parameters $\varepsilon_t$ and $\mu_t$ (also local). For
$\varepsilon_t$ and $\mu_t$ the closed form relations were derived.
These parameters allow one to match normal components of averaged
electric displacement $<\-D>$ and magnetic flux $<\-B>$ at all
boundaries. This turned out to be possible up to $qa\approx 1$ and
for \emph{all angles of incidence} \cite{EuPJ1}. The last fact means
that MP of both central and transition layers at fixed frequency do
not depend on the direction of propagation. It is an obvious feature
of local MP. Using this model the theoretical reflection coefficient
for arbitrary number $N$ of lattice unit cells across the layer
(starting from $N=1$) turned out to be practically equal to that
obtained by exact simulations \cite{EuPJ1,Sim}.

Though the use of local MP \cite{SimPRB,Sim,EuPJ,EuPJ1} makes the
problem of the plane-wave reflection in the frequency range
\ref{eq:fb} more difficult these parameters are, at least, as
important as non-local MP. Once extracted, tensors of local MP are
independent on the wave incidence angle and polarization. Non-local
MP at the same frequency should be measured or simulated separately
for all possible directions of the wave vector. Local MP can
describe the interaction between the medium with wave packages,
moreover, with evanescent waves. Non-local MP are applicable for the
only propagating wave. For the only wave they do not violate the
passivity condition in the effective medium. In a plane wave
electric and magnetic fields are related through the wave impedance.
Then the negative electric or magnetic losses (wrong sign of
$\varepsilon_{\rm eff}$ for magnetic MTM \cite{Smith}--\cite{Yen}
or, vice versa, wrong sign of $\mu_{\rm eff}$ for dielectric MTM)
are compensated by the positive magnetic or electric losses. A
similar speculation can be done on the causality. For wave packages
and for evanescent waves it is not so because in this case the
electric field and magnetic field can be locally decoupled. For
example, in the near zone of a wire antenna its electric field
dominates. Then the wrong sign of electric losses in the substrate
implies the energy generation, and the effective medium turns out to
be active. It is also possible to show the violation of the
causality in these cases. So, non-local MP give not enough insight
on MTM, and to know local MP is important.

In the present paper it will be shown how to find local constitutive
parameters through $R$ and $T$ of a slab.
However, before it, let us understand why MP obtained in
\cite{Smith}--\cite{Yen} and other similar works are non-local.
To answer this question one can discuss the meaning of non-local MP
in terms of the transmission-line theory. Consider a problem of the
plane-wave reflection by a semi-infinite lattice of dipole
scatterers. The strict analytical solution of this problem
was found in \cite{genrefl}. For simplicity let us restrict by the
special case of normal incidence to a orthorhombic lattice of
dipoles located in free space. Also we assume that the frequency
range satisfies to the condition $k_0a=\omega
a\sqrt{\varepsilon_0\mu_0}<\pi$, where $a$ is the lattice period in
the direction of propagation (in this frequency region the only
eigenwave can propagate). Then the reflection coefficient of the
lattice takes form \cite{genrefl}: \e r={\sin{(k_0-q)a\over 2}\over
\sin{(k_0+q)a\over 2}}\Pi. \label{eq:la}\f Here $q$ is the eigenwave
wavenumber and the factor $\Pi$ for which a closed-form expression
was obtained in \cite{genrefl} takes into account all polaritons
excited at the interface. It was proved in \cite{genrefl} that at
the frequencies of our interest this factor has unit absolute value.
So, it can be represented as an imaginary exponential
$\Pi=e^{j\Phi}$. Then the influence of polaritons can be taken in
account by a displacement of the interface to which the new
reflection coefficient $r'$ is referred with respect to the upper
crystal plane to which reflection coefficient $r$ refers in formula
\ref{eq:la}. After it we can rewrite \ref{eq:la} in the form \e
r'=r\exp(-j\Phi)={Z_B-1\over Z_B+1},\quad Z_B= {\tan (k_0a/2)\over
\tan (qa/2)}.
 \label{eq:ZBloch}\f
The phase shift $\Phi$ is rather small for $qa<1$ \cite{genrefl}.

If $max(qa/2,k_0a/2)\ll 1$ the sine and tangent functions in
\ref{eq:la} and \ref{eq:ZBloch} can be replaced by their arguments.
Then we obtain from \ref{eq:ZBloch} $Z_B=k_0/q$ that gives the
approximation of the continuous medium. Really, if we compare
$Z_B=k_0/q=1/\sqrt{\varepsilon_{\rm eff}\mu_{\rm eff}}$ with
\ref{eq:standard} we obtain $\mu_{\rm eff}=1$,
$q=k\sqrt{\varepsilon_{\rm eff}}$ and \ref{eq:ZBloch} becomes the
Fresnel formula for the reflection by a continuous dielectric
half-space. The non-locality of the effective permittivity extracted
through $R$ in this case is negligible. The practical condition of
applicability of this approximation follows from numerical
properties of tangent function and reads as \e
max\left({a\over\lambda},{a\over\lambda_{\rm eff}}\right)<
0.01.\label{eq:fb1}\f This condition gives the upper frequency bound
up to which formulas \ref{eq:rt} and \ref{eq:standard} are
compatible with approximation of the continuous medium. The
frequency region \ref{eq:fb1} is the same for which transition
layers in the theory \cite{Sim} have no visible influence. For
practical implementations of MTM the frequencies satisfying
\ref{eq:fb1} are not interesting.

At frequencies \ref{eq:fb} the normalized impedance
$Z=\sqrt{\varepsilon_{\rm eff}/\mu_{\rm eff}}$ can be understood
through parameters of a periodically loaded transmission line (TL).
Formula \ref{eq:ZBloch} obtained as the \emph{strict} solution of
the boundary lattice problem is known in the theory of periodically
loaded TL. It represents the so-called \emph{Bloch impedance} of a
line loaded by shunt lumped impedances with period $a$ (formula
(5.117) from \cite{Tretyakov}). The Bloch impedance can be defined
as the characteristic impedance of the homogenized periodically
loaded line \cite{ Elefter} normalized to the impedance $\eta$ of
the host matrix. Notice, that the dispersion equation of a TL with
shunt loads is known (e.g. \cite{Pozar}): \e \cos(qa) = \cos(ka)
+\frac{j}{2Z_{\rm load}}\sin(ka), \label{eq:qY} \f where $k$ is the
matrix wavenumber and $Z_{\rm load}$ is the normalized impedance of
loads. The equation \ref{eq:qY} coincides with equation (5.226) from
\cite{Tretyakov}, thoroughly derived  for an infinite lattice of
electric dipoles at frequencies $ka<\pi$. So, the problem of the
reflection and transmission in dipole lattices can be correctly
solved replacing the crystal planes by effective lumped loads and
using the model of a periodically loaded TL. The reflection
coefficient $R$ of a semi-infinite lattice is directly related to
its Bloch impedance $Z=Z_B$. If we measure $R$ and $T$ for a
composite slab with integer number of cells across it, these
parameters will uniquely determine $q$ and $Z$ (and, consequently,
non-local MP). Thus, one can conclude that MP calculated through $R$
and $T$ using \ref{eq:rt} and \ref{eq:standard} are
\emph{transmission-line material parameters} (TLMP), introduced for
TL networks in 2002 in works \cite{Elefter,Itoh}. The idea of TLMP
can be better understood from the comparison of Figs. \ref{MM} and
\ref{figa}.

To understand the reason why TLMP are non-local let us show that
$Z_B$ can be expressed in terms of the Bloch expansion for
microscopic fields in lattices. Namely, if a linearly polarized
eigenmode with wave vector $\-q$ propagates along the axis $z$ in a
lattice with periods $(b_x,b_y,b_z=a)$ the Bloch impedance is equal
to the ratio of zero-order spatial harmonics denoted as $\-E_0$ and
$\-H_0$ in expansions: \e \left\lbrace\begin{matrix} \-E(\-r)\
\\\-H(\-r)\end{matrix}\right.=\left\lbrace\begin{matrix}\-E_0\
\\\-H_0\end{matrix}\right.e^{-jqz}+\sum_{\bf n\ne
0}\-E_{\-n} e^{-j(qz+\-G_{\bf n}.\-r)}, \label{eq:the}\f where
$\-G_{\bf n}=\left({2\pi n_x/b_x},{2\pi n_y/b_y},{2\pi n_z\over
a}\right)$ are multiples of generic lattice vector. From
\ref{eq:the} we have \e \left\lbrace\begin{matrix} \-E_0\
\\ \-H_0\end{matrix}\right.={1\over a}\int\limits_{-a\over 2}^{a\over 2}\left\lbrace\begin{matrix} \-E_{TA}(z)\
\\ \-H_{TA}(z)\end{matrix}\right.
e^{+jqz}dz, \label{eq:EAC}\f where the index $TA$ means the
transverse averaging that allows one to get rid of $G_x,\ G_y$ in
\ref{eq:the}. It is evident that these integrals are exactly equal
to $E_{TA}(-a/2)=E_{TA}(a/2)\exp(jqa)$, $
H_{TA}(-a/2)=H_{TA}(a/2)\exp(jqa)$. The ratio $Z_B=E_{TA}(\pm
a/2)/\eta H_{TA}(\pm a/2)$ can be interpreted as that of effective
voltage and effective current taken at the input or output of the
cell. It is an alternative definition of the normalized Bloch
impedance of periodically loaded TL with unit cell length $a$
\cite{Caloz}. The Bloch impedance that determines (together with
$q$) non-local MP of a lattice is equal to: (a) the characteristic
impedance of the equivalent homogenized TL, (b) the ratio of
fundamental Bloch harmonics of electric and magnetic fields. The
last fact is the reason of the non-locality of TLMP.

Really, in the definition of local MP we deal with averaged electric
and magnetic fields $<\-E>$, $<\-H>$ which are simple integrals of
the microscopic field over the domain around the observation point.
Integrating Maxwell's equations for microscopic fields does not
change these equations and they hold for $<\-E>$ and $<\-H>$ and
averaged polarizations. However, they do not hold for zeroth Bloch
harmonics of microscopic fields taken separately. All high-order
terms of expansion \ref{eq:EAC} give a nonzero contribution into
$<\-E>$ and $<\-H>$! Treating $\-E_0$ and $\-H_0$ substituted into
Maxwell's equations as averaged fields leads to errors which can be
interpreted as \emph{fictitious magnetization of p-lattices} and
\emph{fictitious polarization of m-lattices}. These fictitious
lattice responses have no physical meaning, consequently, one should
not expect the locality of TLMP.

\section{Extraction of local material parameters}

\begin{figure}[btp]
\begin{center}
\includegraphics[width=7.5cm]{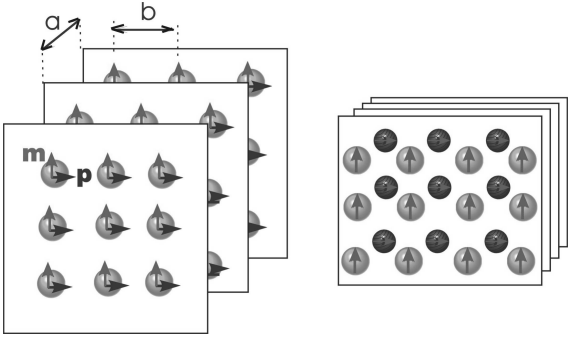}
\caption{Presentation of a p-m-lattice as a set of dipole crystal
planes. Left: every particle has both electric and magnetic moments.
Right: electric and magnetic particles are different.} \label{MM}
\end{center}
\end{figure}

In order to explain the extraction of local MP from $R$ and $T$ of a
MTM slab, let us, first, consider an infinite lattice of electric
and magnetic dipole particles. Assume, that a plane wave with wave
vector $\-q=q\-z_0$ propagates in an infinite lattice shown in Fig.
\ref{MM}. The lattice is formed by orthorhombic unit cells of sizes
$b\x b\x a$ (for simplicity one can skip the case of the anisotropy
in the $(x-y)$plane which gives no effect for the $z$-propagation).
\begin{figure}[btp]
\begin{center}
\includegraphics[width=7.5cm]{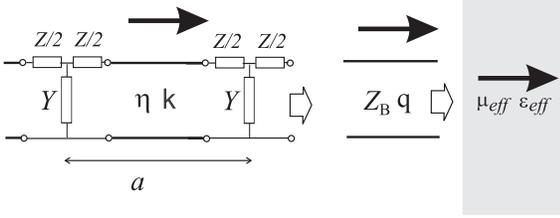}
\caption{Presentation of the p-m-lattice shown in Fig. \ref{figa} as
(1) periodically loaded transmission line, (2) homogenized
transmission line with Bloch impedance $Z_B$ and wavenumber $q$ ($k$
and $\eta$ are unloaded wavenumber and characteristic impedance,
respectively), and (3) effective magneto-dielectric medium with
non-local MP $\varepsilon_{eff}$ and $\mu_{eff}$.} \label{figa}
\end{center}
\end{figure}

In the left panel of Fig. \ref{MM} the particles are assumed
possessing both electric and magnetic moments. In the right panel
the electric and magnetic scatterers are different particles. The
formulas of the local field method are the same for both these
geometries. This is so since the quasistatic interaction between
electric ($\-p$) and magnetic ($\-m$) dipoles is absent in both
these structures (any p-dipole is not affected by m-dipoles lying in
the same crystal plane and vice versa). Let a reference
($0$-numbered) p-m-particle located at the origin have electric and
magnetic polarizabilities $a_{ee},\ a_{mm}$. These polarizabilities
can be resonant in the same frequency range. Let $E^{\rm loc}$ and
$H^{\rm loc}$ denote the $x$-component of the local electric field
and $y$-component of the local magnetic field, respectively. Then
the p-dipole and m-dipole of the particle can be expressed as: \e
p_0 = a_{ee} E^{\rm loc},\quad m_0  =   a_{mm} H^{\rm loc}.
\label{eq:pm} \f Local fields are produced by other particles
located at points $(x=n_xb,y=n_yb,z=n_za)$) and can be expressed
through $p_0$ and $m_0$ since $p(\-n) = p_0\exp({-jn_zqa})$ and
$m(\-n)  = m_0\exp({-jn_zqa})$: \e E^{\rm loc} =
C_{ee}p_0+C_{em}m_0,\quad H^{\rm loc} = C_{mm} m_0+C_{me}p_0.
\label{eq:loc}\f  Substituting \ref{eq:loc} into \ref{eq:pm} we come
to a system
\begin{eqnarray}
{1\over a_{ee}}p_0 &=& C_{ee}p_0+C_{em}m_0,
\label{eq:ploc1} \\
{1\over a_{mm}}m_0 &=& C_{mm}m_0+C_{me}p_0, \label{eq:mloc1}
\end{eqnarray}
A closed-form expression for cross-coupled interaction factor
$C_{em}=C_{me}$ was derived in \cite{Simovski}. For
$C_{ee}=C_{mm}/\eta^2$ one can find in \cite{Tretyakov} the
expression: \e C_{ee}={\eta\omega\over 2 b^2}\left(C_0+C_{NF}+{\sin
ka\over \cos ka - cos q a} \right)+j{k^3 \over 6 \pi \epsilon_m},
\label{eq:exact}\f where $k=\omega\sqrt{\mu_m\varepsilon_m}$ is the
host matrix wavenumber and $\eta=\sqrt{\mu_m/\varepsilon_m}$ is its
wave impedance (here $\varepsilon_m$ and $\mu_m$ are absolute
permittivity and permeability of the matrix). In \ref{eq:exact} the
trigonometric term describes the wave interaction of the reference
p-dipole with p-dipoles of other crystal planes, the real value
$C_{NF}$ describes the near-field interaction with p-dipoles of
other crystal planes. The remain of \ref{eq:exact} is the real value
$C_0$ which is responsible for the near-field interaction of the
reference p-dipole with other p-dipoles of the reference crystal
plane. For $C_0$ a simple expression was derived in
\cite{Tretyakov}:
$$
C_0={1\over 2}\left({\cos kbs\over kbs}-\sin kbs\right),
$$
where $s \approx 0.6954$. At very low frequencies $kb< 0.5$ the
approximation $C_0\approx 0.36$ is quite accurate \cite{Tretyakov}.

Equating the determinant of \ref{eq:ploc1}, \ref{eq:mloc1} to zero
and neglecting $C_{NF}$ (as it was shown in \cite{Tretyakov}, it is
a suitable approximation for the case $a\le b$) we obtain the
dispersion equation (an alternative equivalent form of it was
derived in \cite{Simovski}):
$$ \cos qa=\cos k a- \sin k a \left({G\over 4}+ {X\over 4}\right)-
$$\e \sqrt{\sin^2 k a \left({G\over 4}- {X\over 4}\right)^2+{G X\over
4}\sin ^2 q a }. \label{eq:general}\f Here the notations were
introduced: \e \left\lbrace\begin{matrix} G\
\\ X\end{matrix}\right\rbrace
={k_0a\over \varepsilon_mV\left(\left\lbrace\begin{matrix} {1\over
a_{ee}}
\\ {1\over a_{mm}}\end{matrix}\right\rbrace
-j{k^3 \over 6 \pi} \left\lbrace\begin{matrix}
{1\over\varepsilon_{m}}
\\ {1\over\mu_{m}}\end{matrix}\right\rbrace\right)-C_0}. \label{eq:TLI}\f
In \ref{eq:TLI} $V=ab^2$ is the unit cell volume. In the lossless
structure the fundamental relations hold for electric and magnetic
polarizabilities \cite{Man,Man1,Sipe}:
\e {\rm Im}{1\over a_{ee}}={k^3\over 6\pi\varepsilon_m},\quad {\rm
Im}{1\over a_{mm}}={k^3\over 6\pi\mu_m}. \label{eq:ManP}\f Then the
parameters $G$ and $X$ in \ref{eq:general} turn out to be
real-valued in lossless lattices.

Consider the averaged fields $<E>$ and $<H>$ defined as \e
<E>(\-r)={1\over V}\int\limits_{V} E_x(\-r+\-r')d^3 \-r',
\label{eq:def1}\f \e <H>(\-r)={1\over V}\int\limits_{V}
H_y(\-r+\-r')d^3 \-r' \label{eq:def2}\f Here $\-r'$ lies within the
unit cell and $E_x,\ H_y$ are components of the microscopic fields.
For microscopic fields Maxwell's equations hold:
$$
{\rm rot}\-E=-j\omega(\mu_m\-H+\-M),\quad {\rm
rot}\-H=j\omega(\varepsilon_m\-E+\-P).
$$
Integrating these equations around the observation point with the
use of \ref{eq:the} we come to Maxwell's equations for averaged
fields and polarizations: \e q<H>=\omega\varepsilon_m<E>+\omega<P>,
\label{eq:fi}\f\e q<E>=\omega\mu_m<H>+ \omega<M>,\label{eq:se}\f
whose solutions take form \e
\left\{\begin{aligned} <E>,\quad <H>\\
<P>,\quad <M>\end{aligned}\right. =
\left\{\begin{aligned} E_L, \quad H_L\\
P_L, \quad M_L\end{aligned}\right.\exp(-jqz). \label{eq:lore}\f The
amplitudes $E_L,\ H_L$ are related with Bloch harmonics $E_n,\ H_n$
of transversely averaged fields as \e
\left\{\begin{aligned} E_L\\
H_L\end{aligned}\right.=\left\{\begin{aligned} E_0/qa\\
H_0/qa\end{aligned}\right.+\sum\limits_{n=1}^{\infty}\left\{\begin{aligned} E_n/(qa+2\pi n)\\
H_n(qa+2\pi n)\end{aligned}\right.\f and at the reference particle
center $x=y=z=0$ we have $<P>=p_0/V$ and $<M>=m_0/V$. Equations
\ref{eq:fi} and \ref{eq:se} at $z=0$ are equivalent to relations \e
q^2=\omega^2\varepsilon_{L}\mu_L,\quad Z_L\equiv {<E>(0)\over
<H>(0)}=\sqrt{\mu_L\over\varepsilon_L}.\label{eq:LLA}\f Here
coefficients $\varepsilon_{L}$ and $\ \mu_{L}$ are defined by
formulas \e
\left\{\begin{aligned} <P>(0)\\
<M>(0)\end{aligned}\right.=\left\{\begin{aligned} (\varepsilon_L-\varepsilon_0)<E>(0)\\
(\mu_L -\mu_0)<H>(0)\end{aligned}\right. \label{eq:def}\f and have
physical meaning of absolute permittivity and permeability. They
should be \emph{local} even at \emph{moderately} low frequencies
\ref{eq:fb} since relate \emph{averaged} fields and \emph{averaged}
polarizations instead of zero-order Bloch harmonics. It is difficult
to prove strictly this fact (see also the discussion in
\cite{Simovski1}), but the explicit examples confirm it.

Let us assume that we know parameters $G$ and $X$ describing the
crystal plane response. The evident way to find local MP is to
express polarizabilities $a_{ee},\ a_{mm}$ through $G,\ X$ using
formulas \ref{eq:TLI} and then to apply the well-known
Clausius-Mossotti formula expressing $\varepsilon_L$ via $a_{ee}$
and $\mu_L$ via $a_{mm}$. However, this formula is quasi-static and
is hardly accurate at moderate frequencies. Formulas \ref{eq:LLA}
show another (not static) way for obtaining local MP through $G$ and
$X$. Really, from \ref{eq:ploc1}, \ref{eq:mloc1} we have
$$
\eta p_0\left({2\over G} -{\sin ka\over \cos qa -\cos ka}
\right)=m_0{\sin qa\over \cos qa -\cos ka},
$$
$$
m_0\left({2\over X} -{\sin ka\over \cos qa -\cos ka} \right)=\eta
p_0{\sin qa\over \cos qa -\cos ka}.
$$
These relations give the auxiliary coefficient one needs in order to
find $Z_L$:
$$\gamma={\eta p_0\over m_0}={\eta <P>(0)\over
<M>(0)}=$$\e\sqrt{G\over X}\sqrt{\cos qa- \cos ka -{G\sin ka\over
2}\over \cos qa- \cos ka -{X\sin ka\over 2}}. \label{eq:gamma}\f
From \ref{eq:def} we obtain (see also in \cite{Simovski,Simovski1}:
$$\gamma=\eta Z_L{(\varepsilon_{L}-\varepsilon_0)\over (\mu_{L}-\mu_0)}.$$
However, from \ref{eq:LLA} one has \e\varepsilon_L={q\over\omega
Z_L},\quad \mu_L={q\over \omega} Z_L.\label{eq:LA}\f Therefore the
wave impedance $Z_L$
can be expressed as \e {Z_L\over \eta}={\gamma k+q\over \gamma q+
k}.\label{eq:ZZZ}\f So, from $G,\ X$ one can find $q$, then
$\gamma$, then $Z_L$, and finally local MP.

Practically, what is suggested is the generalization of the
quasistatic approach based on the Clausius-Mossotti formula to the
range of moderate frequencies. In the quasistatic limit our result
for $\varepsilon_L$ and $\mu_L$ must be the same as this classical
result. Let us prove it for simplicity for a cubic p-lattice. In
this case $X=0$ and \ref{eq:general} simplifies to
\begin{equation}
 \cos(qa) = \cos(ka) -\frac{G}{2}\sin(ka).
 \label{eq:qY}
\end{equation}
Substitution of $\gamma\rightarrow\infty$ into \ref{eq:ZZZ} gives
$Z_L=\eta k/q$ and $q=k_0\sqrt{\varepsilon_L}$. For low frequencies
($ka\ll \pi$) we take in account two terms of the Taylor expansion
of trigonometric functions in \ref{eq:qY} and substituting
expression \ref{eq:TLI} for $G$ obtain:
$$
1-{(k_0a)^2\varepsilon_L\over
2\varepsilon_0}=1-{(k_0a)^2\varepsilon_m\over
2\varepsilon_0}-{(k_0a)^2\varepsilon_m\left(1-{(k_0a)^2\varepsilon_m\over
6\varepsilon_0}\right)\over 2\left[ V\varepsilon_m{\rm
Re}\left({1\over a_{ee}}\right)-C_0\right]}.
$$
Neglecting the small term $(ka)^2/6\ll 1$ we obtain the
Clausius-Mossotti equation for local permittivity:
$$
\varepsilon_L=\varepsilon_m+{1\over V{\rm Re}\left({1\over
a_{ee}}\right)-{C_0\over \varepsilon_m}}.
$$
Our factor $C_0=0.36$ differs slightly from classical $0.33$. This
difference is related with the ignorance of near-field interaction
between adjacent crystal planes. The needed correction can be easily
introduced in the theory. Taking in account three terms in the
Taylor expansion of cosine and sine functions gives the frequency
corrections of the order $(k_0a)^2$ and $(k_0a)^3$ to the
Clausius-Mossotti relation. These corrections were obtained in
another way in work \cite{Sim}. A similar proof can be done for
$p-m-$lattices where one comes in the limit case $(ka,qa)\ll \pi$ to
the Clausius-Mossotti formulas for both $\varepsilon_L$ and $\mu_L$.

Now, let us relate $\varepsilon_L,\ \mu_L$ with $R$ and $T$
coefficients of a finite slab.  The simplest way is to find $q$ and
$Z=Z_B$ entering \ref{eq:standard} inverting \ref{eq:rt} in a usual
way and then find $G$ and $X$ through them. With $G,\ X$ we can
calculate local MP. Let us then relate Bloch impedance $Z_B$ with
$G$ and $X$. For it one can apply the transfer matrix method. For
the case under consideration (the normal propagation with respect to
$(x-y)$ crystal planes) the transfer matrix of a lattice unit cell
$\mathbf{F}$ with components $(A,B,C,D)$ is defined as
\begin{equation}
\label{ABCD}
    \begin{bmatrix} E_{TA}\left(-{a\over 2}\right)/\eta \\H_{TA}\left({a\over 2}\right)  \end{bmatrix}=
    \begin{bmatrix}
        A & B\\
        C & D
    \end{bmatrix}
    \begin{bmatrix} E_{TA}\left({a\over 2}\right)/\eta \\H_{TA}\left({a\over 2}\right) \end{bmatrix}.
\end{equation}
It can be calculated as a product of transfer matrices of its three
portions: two $a/2$-long pieces of the TL modeling the host medium
and the lumped circuit modeling the p-m-crystal plane (see Fig.
\ref{figa}):
\begin{equation}
 \mathbf{F}= \mathbf{F}_{TL} \mathbf{F}_{YZ} \mathbf{F}_{TL},
\label{eq:AOY}
\end{equation}
Here
\begin{equation}
 \mathbf{F}_{TL} =
    \begin{bmatrix}
        \cos ({ka\over 2}) & j\sin ({k a\over2})\\
        j\sin ({ka\over2}) & -\cos ({k a\over2})
    \end{bmatrix},\quad
 \mathbf{F}_{YZ} =
    \begin{bmatrix}
        1 & Z \\
        Y & -1
    \end{bmatrix}.
\end{equation}
Values $G=-jY$ and $X=-jZ$  are parameters we introduced above as
$G$ and $X$ and are given by \ref{eq:TLI}. For a lossless structure
$G$ and $X$ are equal to the shunt imittance and series reactance of
the loading, respectively. Parameters $Y$ and $Z$ are known in the
theory of planar grids as the grid electric admittance and the grid
magnetic impedance and describe the ratio between tangential
electric and magnetic fields in the grid plane and electric and
magnetic surface currents induced in the grid (all averaged over the
grid periods). Even if particles have sizes comparable to the period
$a$ one can attribute effective currents to the crystal plane $z=0$.
This procedure is basically the same as replacing the finite
particles by point p- and m-dipoles which is an excellent
approximation for many practical cases.

The Bloch impedance is given by relation \cite{Pozar}: \e
  Z_B = \frac{B}{\sin qa}
. \label{eq:ZZ} \f After some algebra we obtain from
\ref{eq:AOY}-\ref{eq:ZZ}: \e Z_B= {\sin qa\over
G\cos^2(ka/2)-X\sin^2(ka/2)+\sin ka} .
 \label{eq:ZB}\f
If one knows $q$ and $Z$ one can express $G$ and $X$ through them
using equations \ref{eq:general} and \ref{eq:ZB}. Then one finds
local MP through $G$ and $X$ using \ref{eq:LA}, \ref{eq:ZZZ} and
\ref{eq:gamma}.

To conclude this section one can notice that studying the transfer
matrix of a cascade of lattice unit cells it is possible to prove
that TLMP \cite{Caloz,Elefter,Itoh} are not mesoscopic. Being found
from $R$ and $T$ of the layer containing $N$ lattice unit cells
across it these MP can be applied for slabs with arbitrary thickness
(if divisible by $a$). Therefore, one can attribute TLMP found for
finite slabs to infinite lattices. Considering the unit cell
transfer matrix $ \mathbf{F}$ one can also prove that constitutive
parameters suggested in works \cite{Pen,Pen1} and discussed in
\cite{Pendrynew} are equivalent to TLMP.

\section{Results}

\begin{figure}
\begin{center}
\subfigure[][]{\label{Yl}\includegraphics[width=6.5cm]{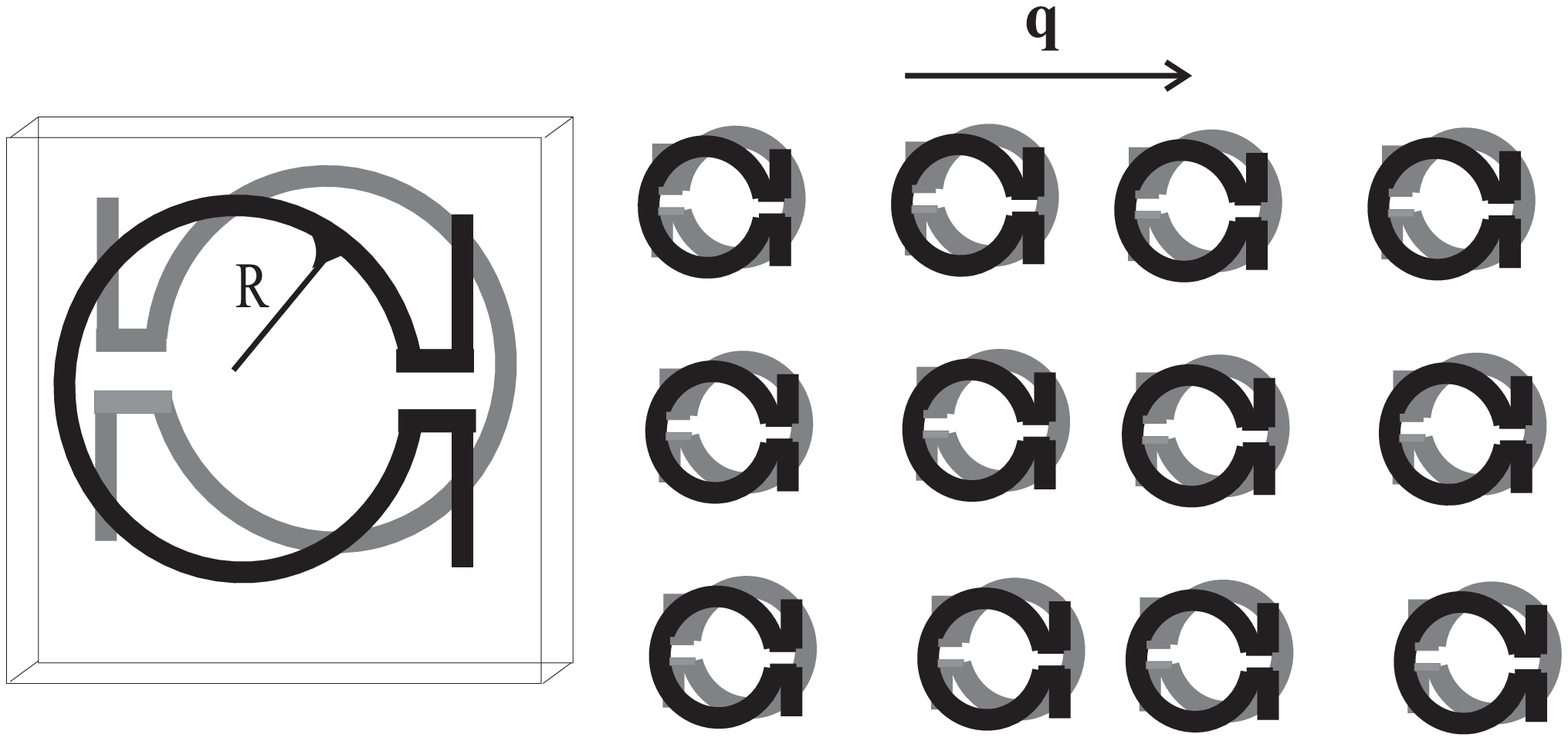}}
\subfigure[][]{\label{Yo}\includegraphics[width=6.5cm]{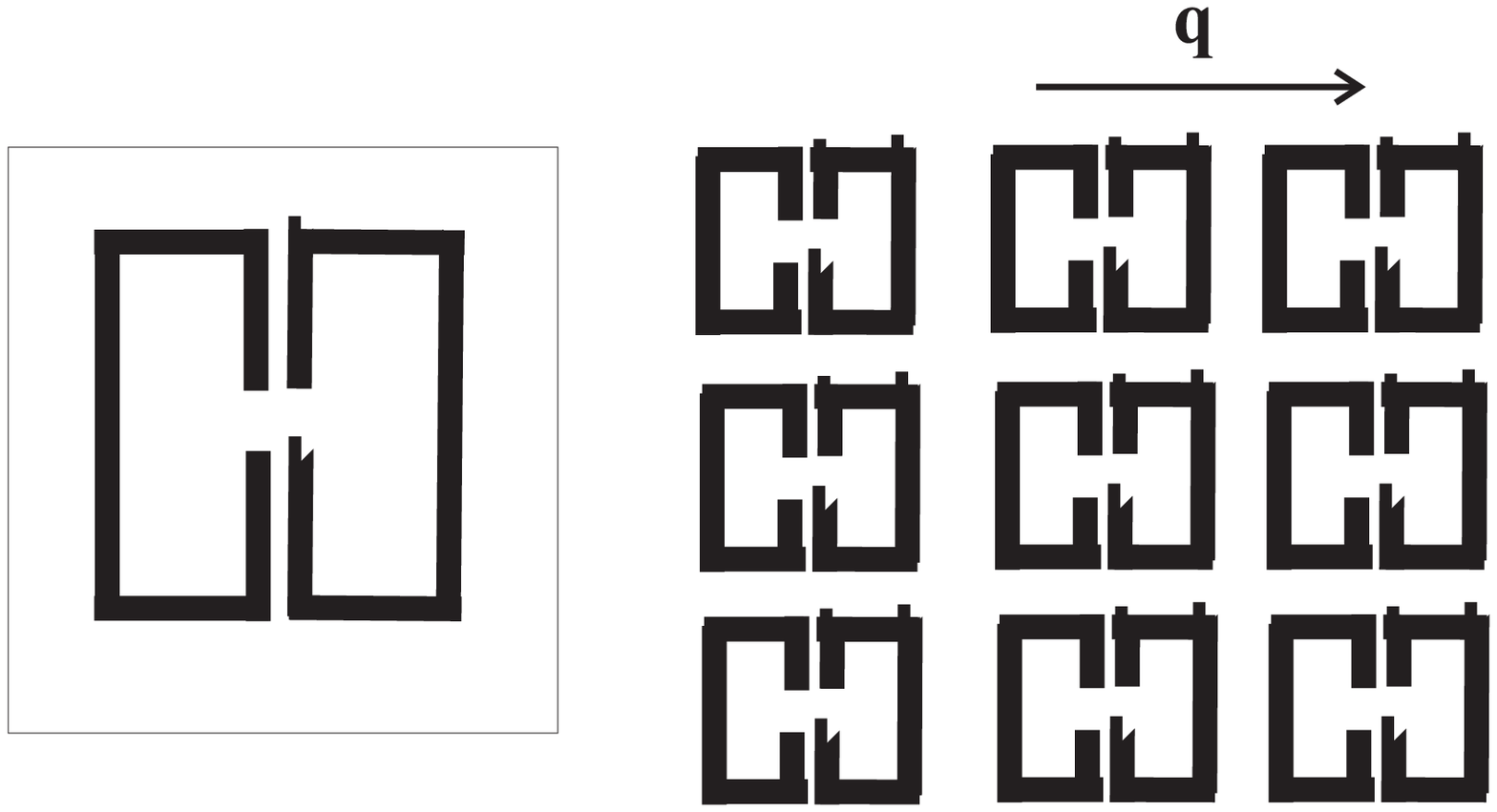}}
\caption{The lattice of pairs of $\Omega$-particles \subref{Yl}, and
the lattice of infrared SRRs suggested by J.B. Pendry and S. O'Brien
in \cite{OBrien0} \subref{Yo}.} \label{figg}
\end{center}
\end{figure}

\begin{figure}
\begin{center}
\subfigure[][]{\label{one}\includegraphics[width=4.2cm]{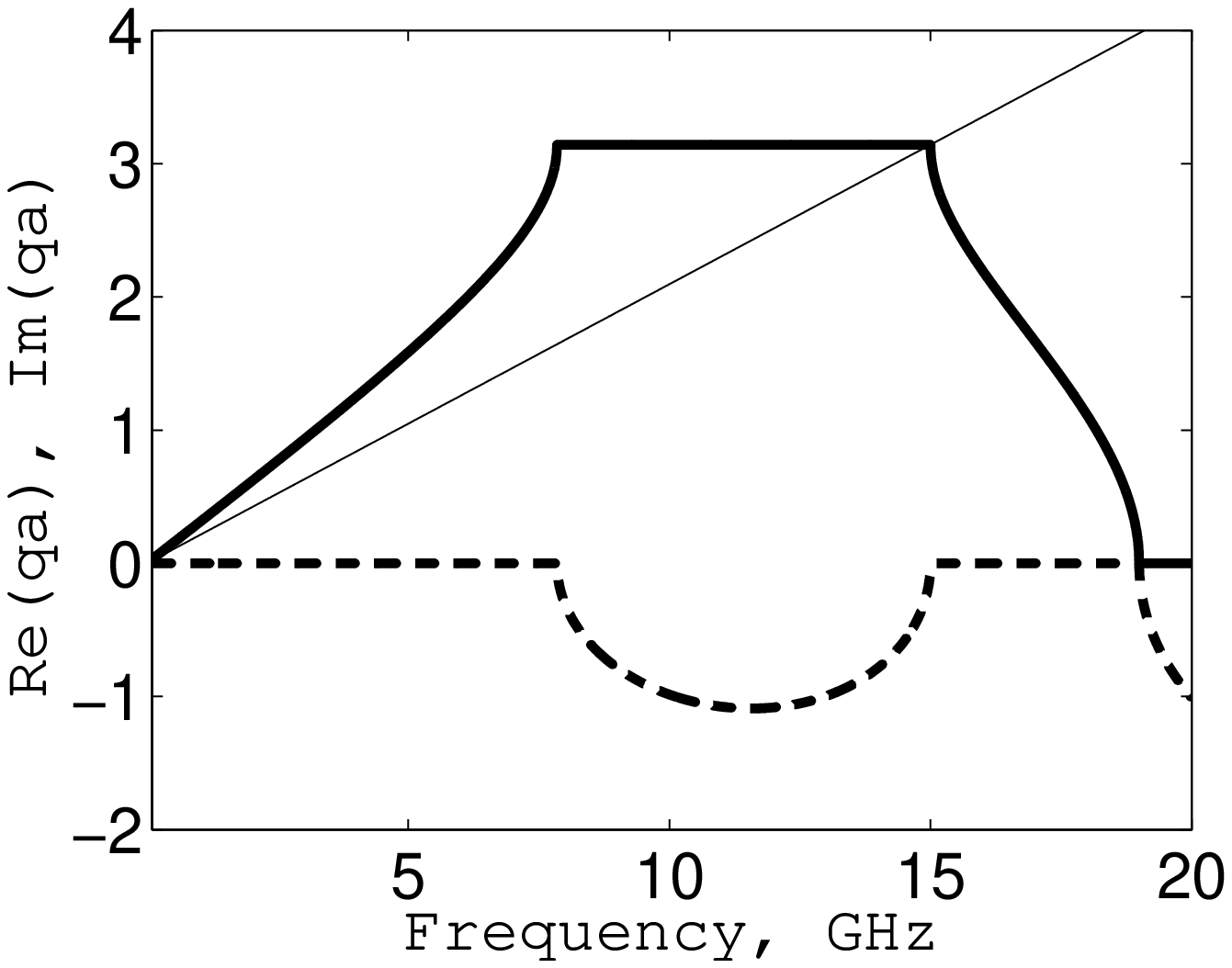}}
\subfigure[][]{\label{two}\includegraphics[width=4.2cm]{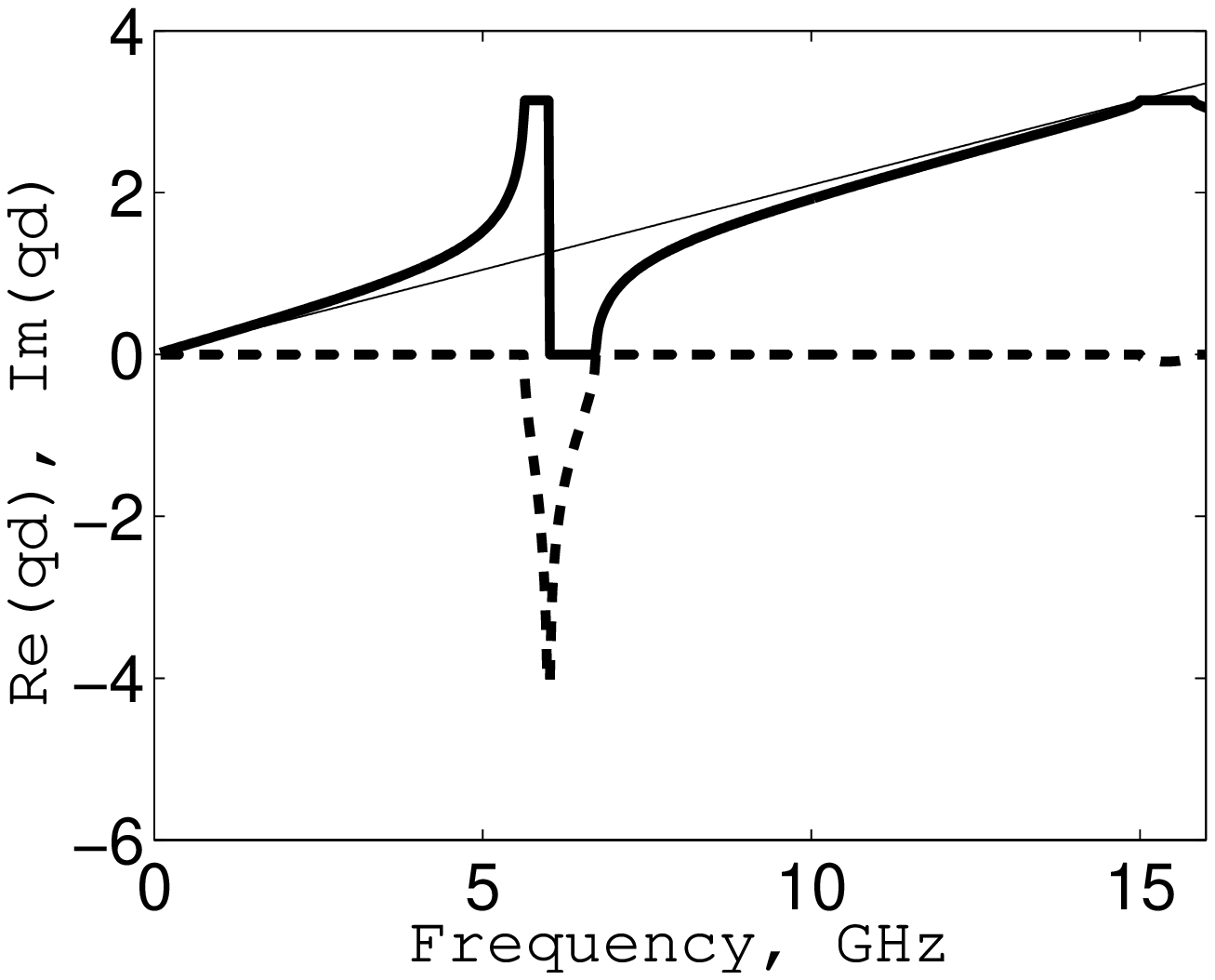}}
\subfigure[][]{\label{Yd}\includegraphics[width=4.2cm]{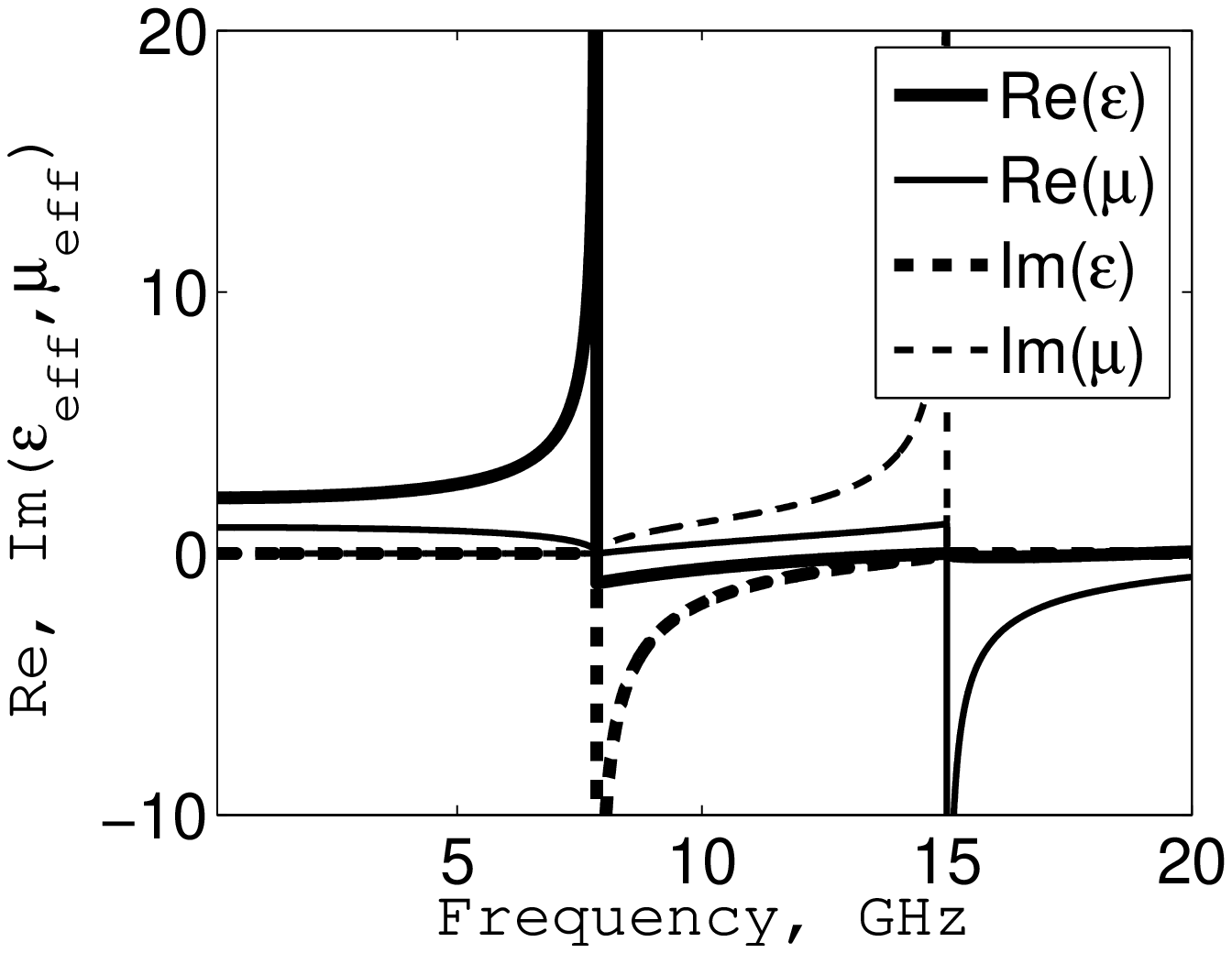}}
\subfigure[][]{\label{YS}\includegraphics[width=4.2cm]{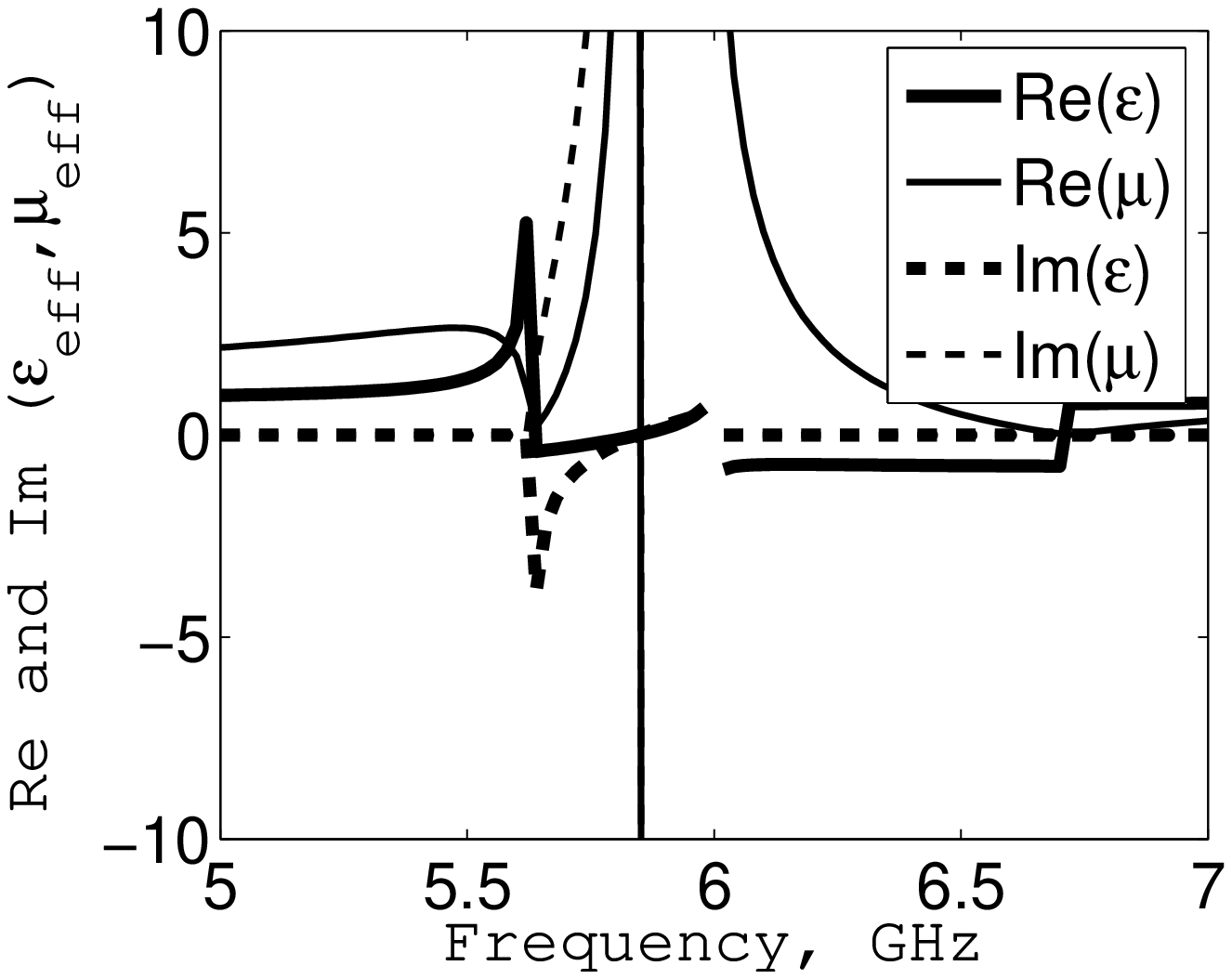}}
\subfigure[][]{\label{YL}\includegraphics[width=4.2cm]{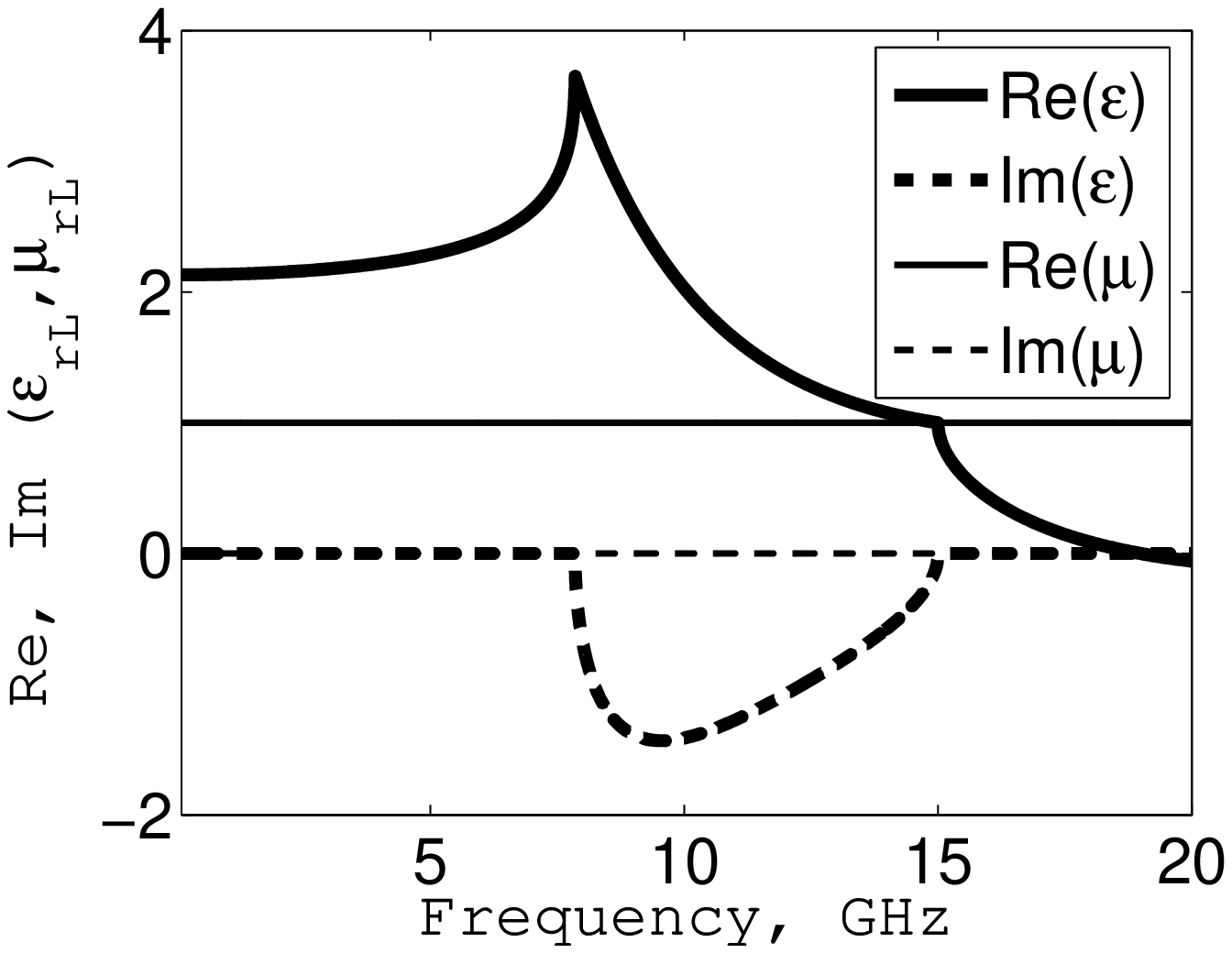}}
\subfigure[][]{\label{YL1}\includegraphics[width=4.2cm]{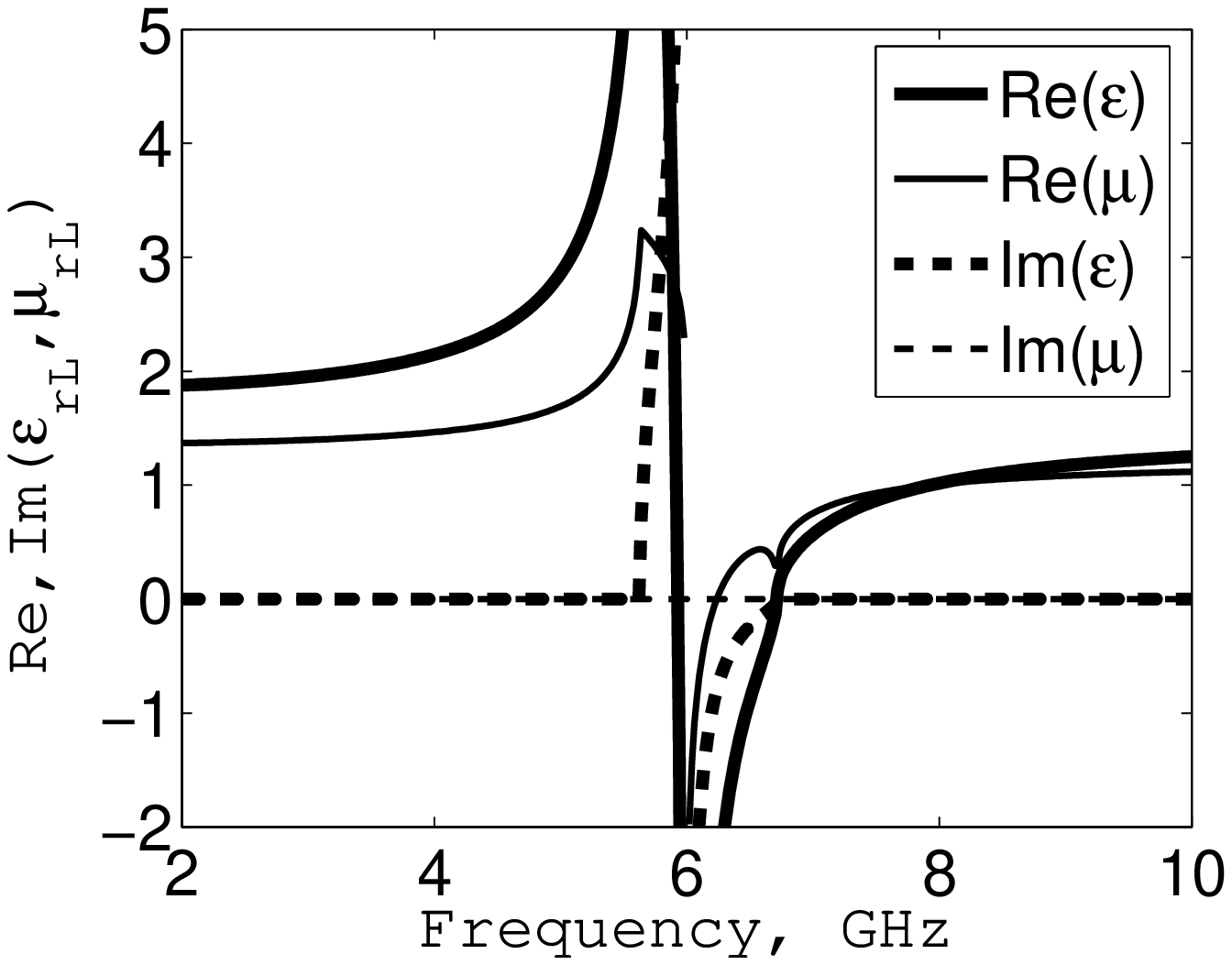}}
\caption{\subref{one}--dispersion in a lattice of non-resonant
dipoles; \subref{two}--dispersion in a lattice of resonant
p-m-particles (pairs of $\Omega$-particles); \subref{Yd}--non-local
MP for the lattice of non-resonant dipoles; \subref{YS}--non-local
MP for p-m-particles; \subref{YL1}--local MP for the lattice of
non-resonant dipoles; \subref{YL1}--local MP for the lattice of
resonant p-m-particles.} \label{figu}
\end{center}
\end{figure}

To illustrate the locality of one set of MP and non-locality of the
other one, two numerical examples of the direct calculation of them
through the known polarizabilities of particles are presented. The
algorithm of this calculation is as follows: \e
(a_{ee},a_{mm})\rightarrow (G,X)\rightarrow
\begin{cases}
(q,Z_B)\rightarrow(\varepsilon_{\rm eff},\mu_{\rm eff})\\
(q,Z_L)\rightarrow(\varepsilon_{L},\mu_{L})
\end{cases}.
\label{eq:one}\f The first arrow implies formulas \ref{eq:TLI}, the
second one implies the solution of \ref{eq:general} and the use of
formula \ref{eq:ZZ} in the upper case and \ref{eq:ZB} in the lower
case. To find local MP formulas \ref{eq:gamma}--\ref{eq:ZZZ} were
used. The first numerical example corresponds to a cubic lattice
formed by capacitively loaded wire dipoles. The second example
corresponds to a lattice of pairs of $\Omega$-particles as is shown
in Fig. \ref{figg} (a).
In the first example the polarizability of particles is nearly
static $a _{ee}\approx l^2C_0$, where $l$ is the effective length of
a wire dipole and $C_0$ is the loading capacitance. However, with
rather large $C_0$ the effect of the presence of non-resonant
dipoles can be significant. This effect is a wide stop-band with
staggered mode near the first lattice resonance as one can see in
Fig. \ref{figu} (a). Notice, that qualitatively same results as
shown in Fig. \ref{figu} (a,c,e) should correspond also to lattices
of metal spheres of radius $2-4$ mm which also behave below 15 GHz
(the lattice period was chosen $a=10$ mm in both examples and the
host matrix was free space) as non-resonant dipoles with rather high
static polarizability.
A lattice of electric dipoles cannot have local magnetic
susceptibility. And the relative local permeability $\mu_{rL}$ is
identically unity in Fig. \ref{figu} (e). The locality conditions
are satisfied until the 1st lattice resonance which occurs at 7.5
GHz. Non-local MP in Fig. \ref{figu} (c) contain non-trivial
permeability and imaginary parts of $\varepsilon_{\rm eff}$ and
$\mu_{\rm eff}$ have opposite signs. Even below 7.5 GHz we observe
in Fig. \ref{figu} (c) the violation of causality.
\begin{figure}
\begin{center}
\subfigure[][]{\label{one}\includegraphics[width=4cm]{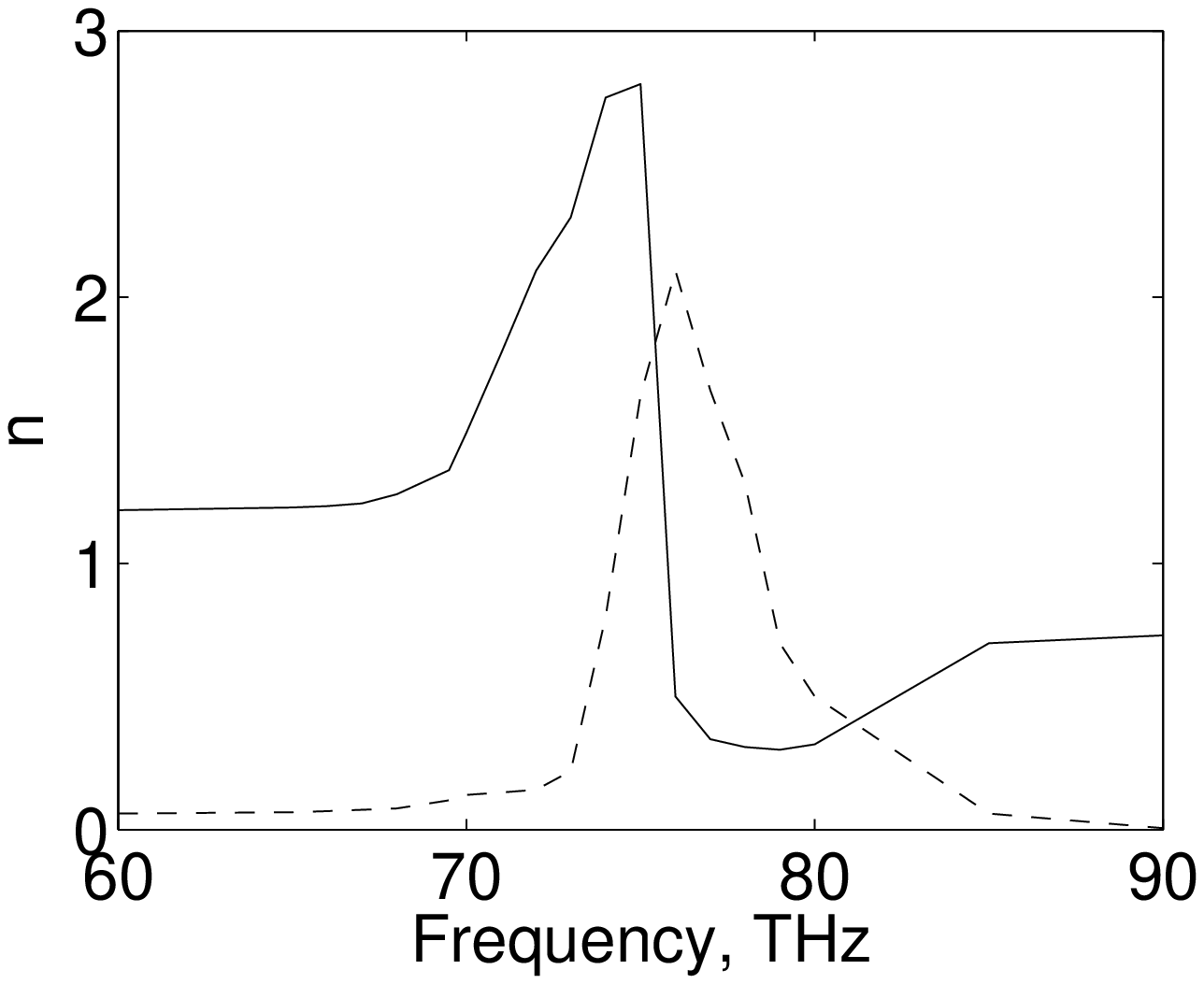}}
\subfigure[][]{\label{two}\includegraphics[width=4cm]{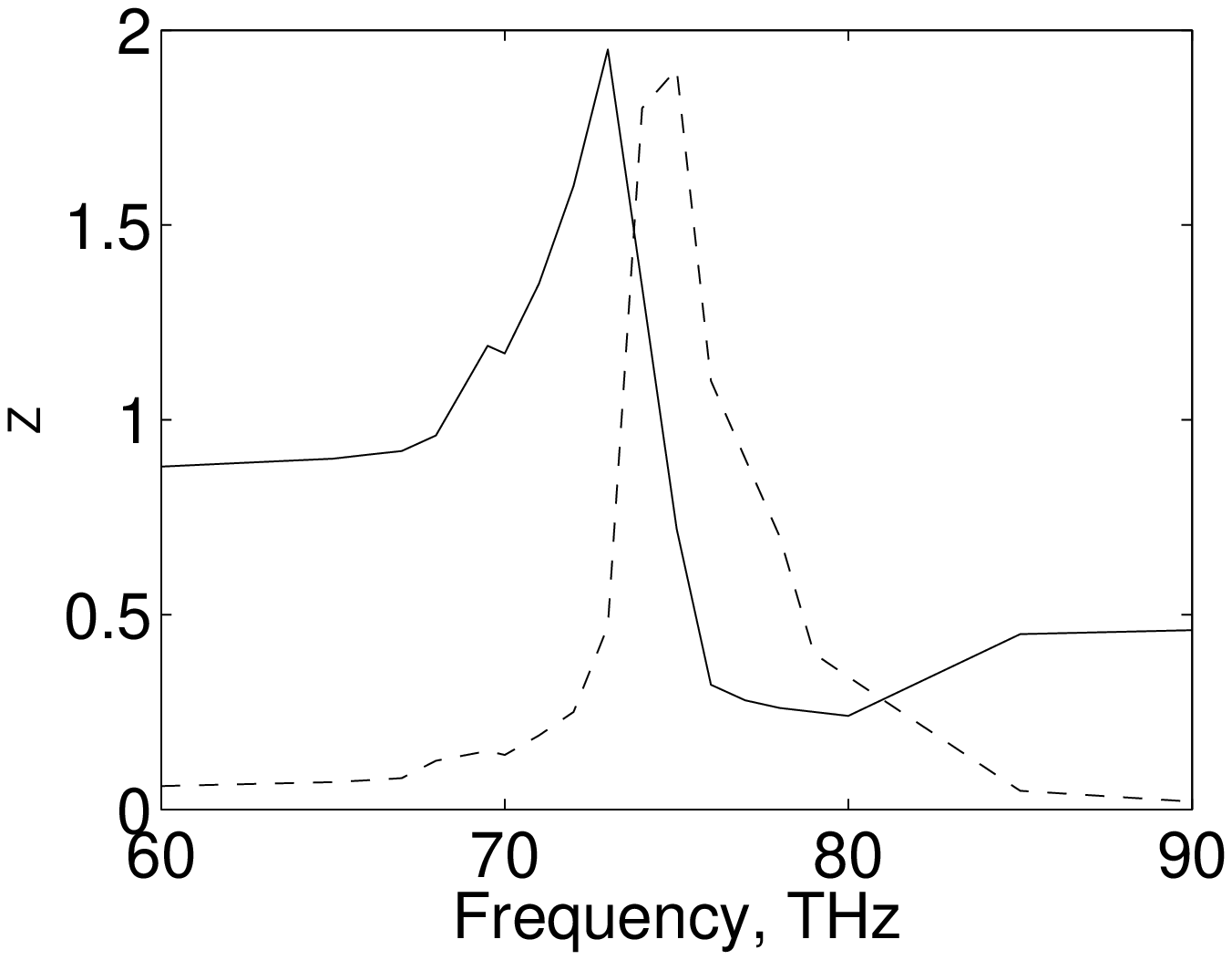}}
\subfigure[][]{\label{tri}\includegraphics[width=4cm]{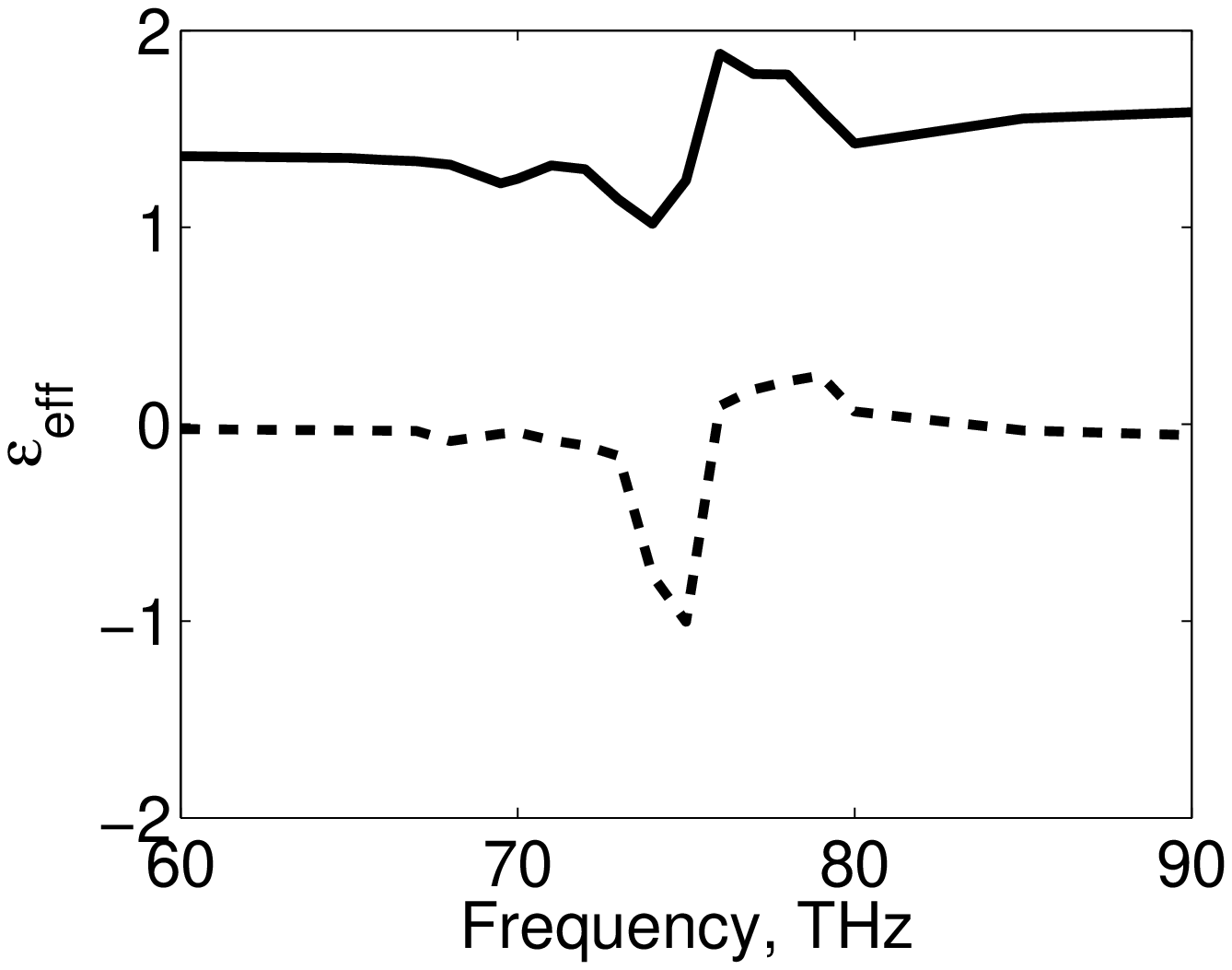}}
\subfigure[][]{\label{ch}\includegraphics[width=4cm]{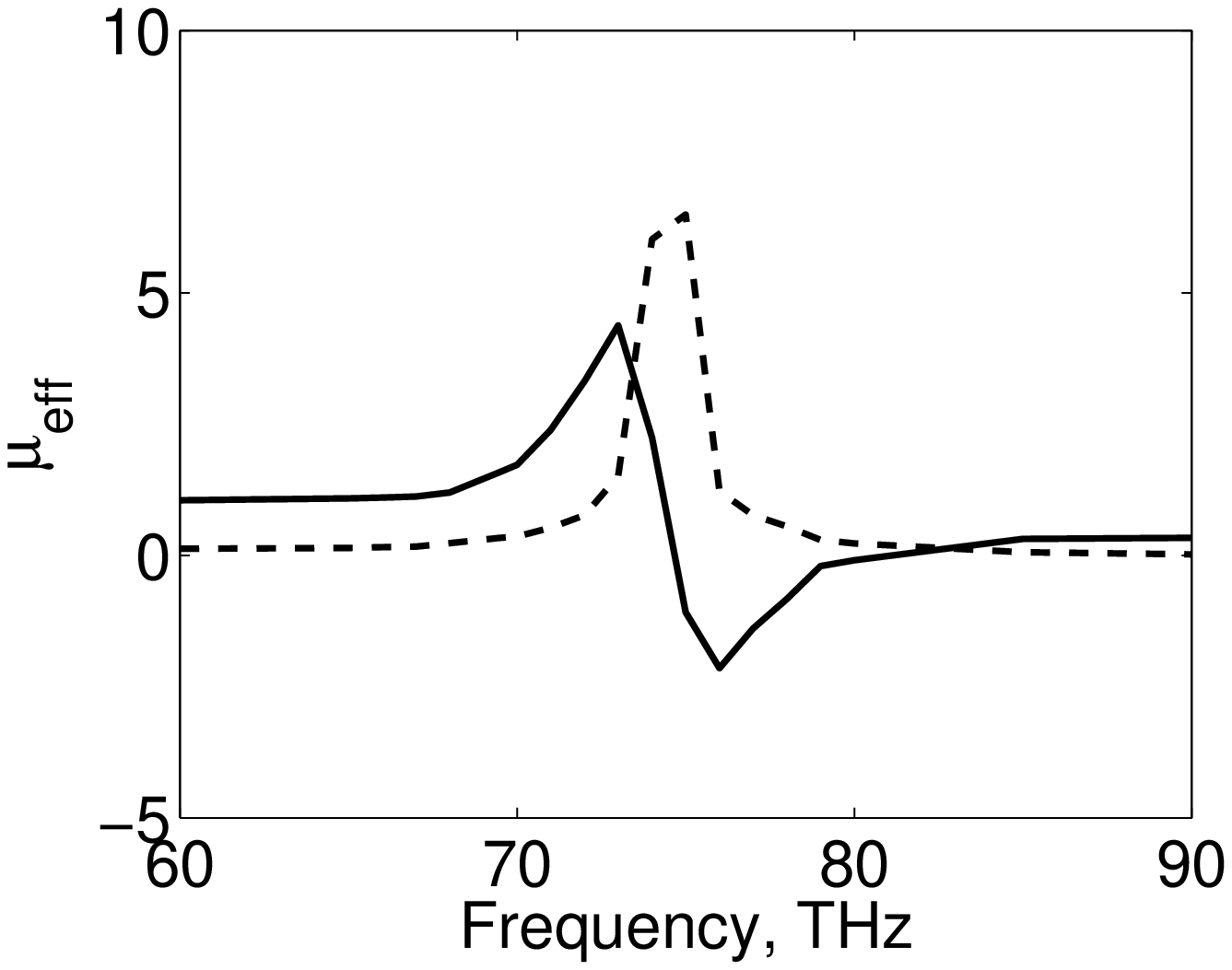}}
\subfigure[][]{\label{p}\includegraphics[width=4cm]{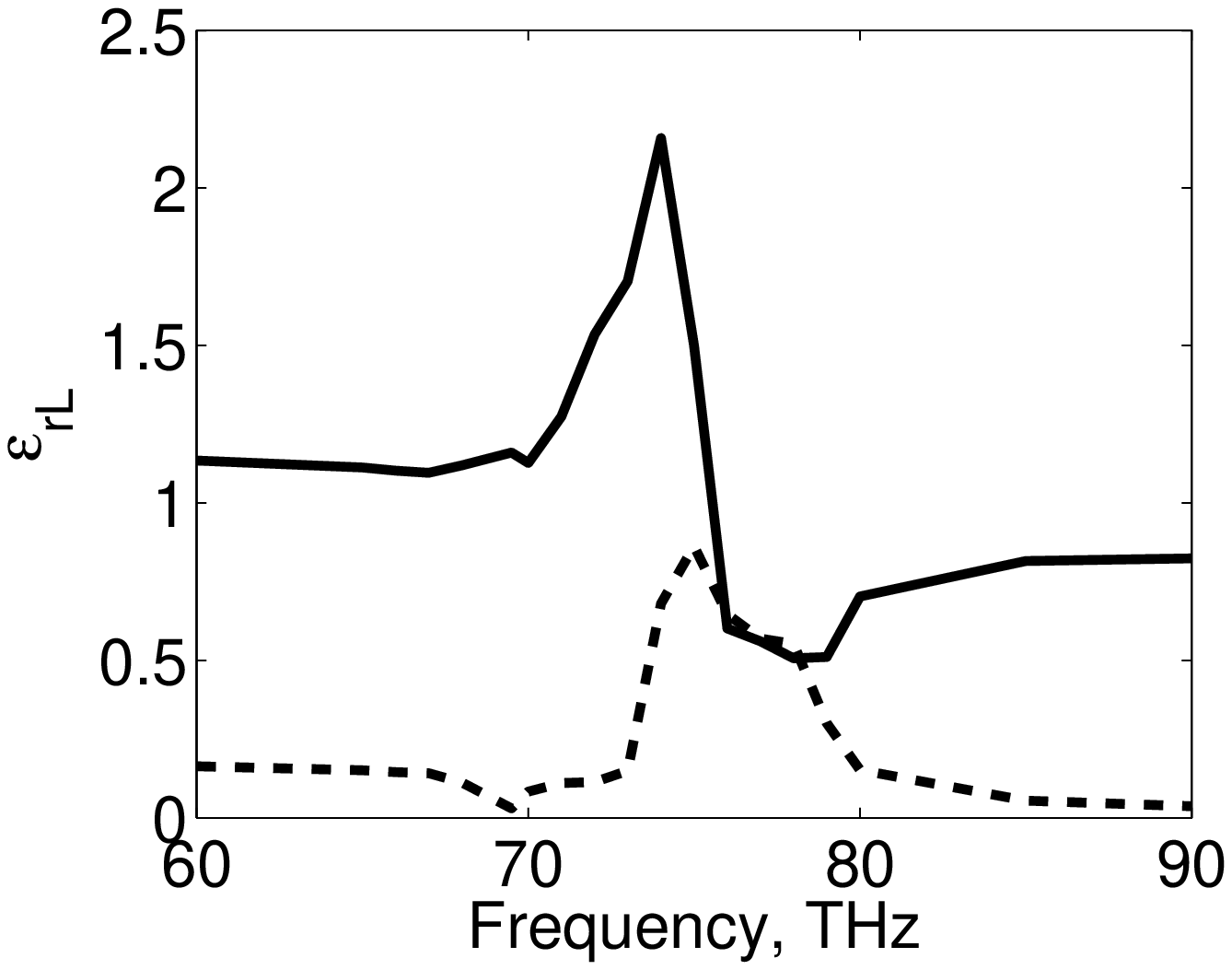}}
\subfigure[][]{\label{sh}\includegraphics[width=4cm]{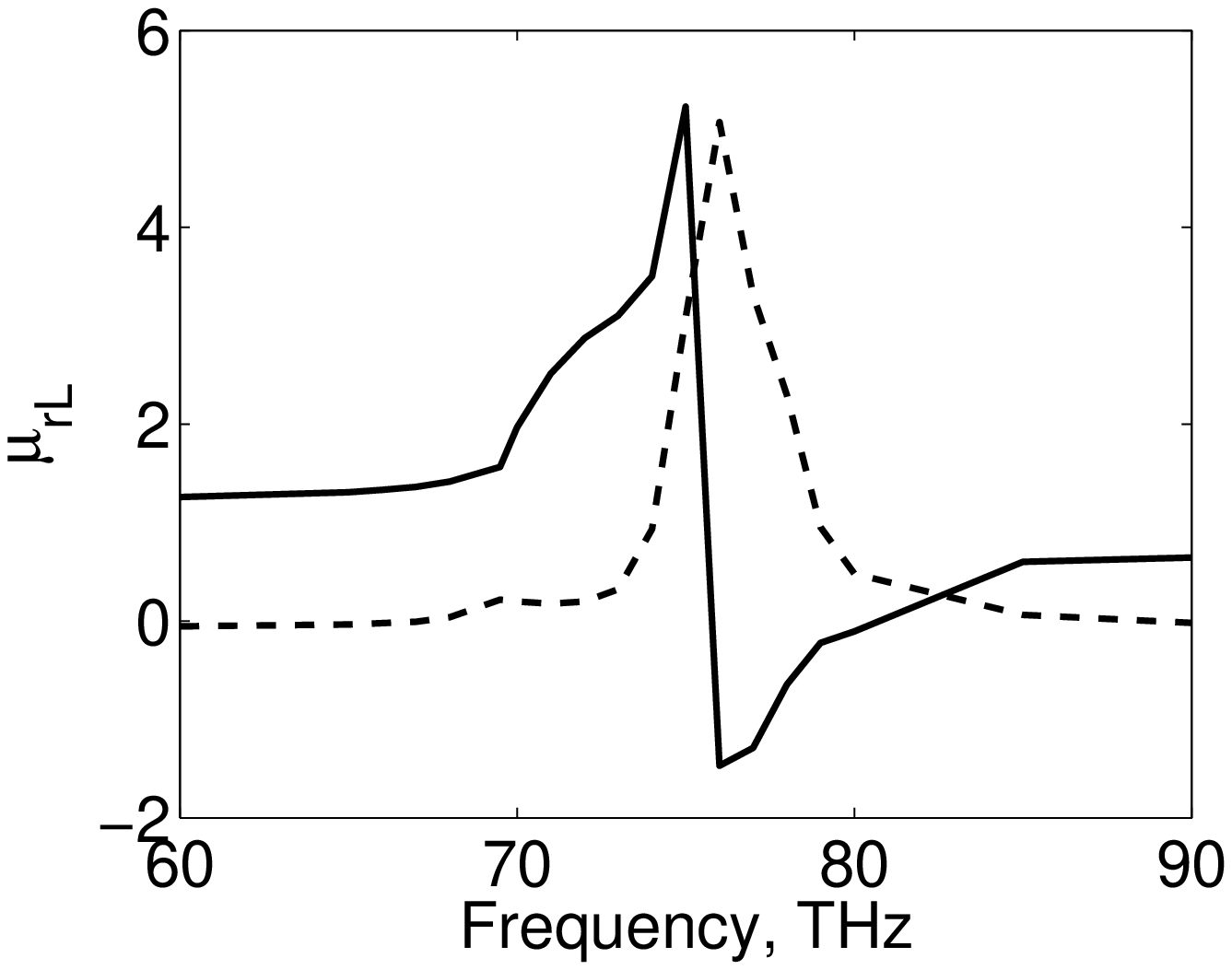}}
\caption{\subref{one}--refraction index extracted in \cite{OBrien2}
from $R,\ T$ coefficients of a slab comprising the lattice of silver
SRRs with period $a=600$ nm; \subref{two}--normalized Bloch
impedance extracted in \cite{OBrien2}; \subref{tri}--non-local
permittivity extracted for this lattice; \subref{ch}--non-local
permeability; \subref{p}--local permittivity; \subref{sh}--local
permeability.} \label{fignew}
\end{center}
\end{figure}

In the second example both electric and magnetic polarizabilities of
doubled $\Omega$-particles are resonant. We calculate them using
formulas \cite{Simovski,Serdyuk}: \e a _{ee}= \frac{A}{{\omega _0^2
- \omega ^2 + j \omega \Gamma}}, ~~~~~~A = {l^2\over
L_0}\label{eq:aee},\f
 \e a_{mm}=\frac{B\omega^2}{{\omega _0^2 - \omega ^2 + j \omega \Gamma}},~~~~~~B =
\frac{{\pi^2\mu_0 ^2 R^4 }}{{L_0 }} \label{eq:amm},\f where $L_0$ is
the inductance of metal rings, $l$ is the effective length of the
electric dipole induced in the particle, $R$ is the ring radius
shown in Fig. \ref{figg}. In this example the p-m-particles resonate
at 6 GHz. Loss factor $\Gamma$ was taken negligibly small in order
to avoid the mess in complex solutions of equation \ref{eq:general}.
The second example is illustrated by plots in Fig. \ref{figu}
(b,d,f). The staggered mode holds in the narrow band $5.8...6$ GHz.
In this band local MP have no physical meaning. Only within this
band the locality of $\varepsilon_{rL},\ \mu_{rL}$ is lost. Outside
it (even at higher frequencies) the sign of ${\rm
Im}(\varepsilon_{rL},\mu_{rL})$ is correct and the frequency
behavior of ${\rm Re}(\varepsilon_{rL},\mu_{rL})$ in the low-loss
region is causal.

In our third example we study the structure for which the input data
are $R$ and $T$ coefficients of a slab. It is filled by a lattice of
silver split-ring resonators (SRRs) shown in Fig. \ref{figg} (b).
This structure operating in the infrared range was studied in
\cite{OBrien0,OBrien2}. For this case the algorithm is as follows:
\e (R,T)\rightarrow (q,Z_B)\rightarrow (G,X)\rightarrow
(\varepsilon_{L},\mu_{L}). \label{eq:two}\f In algorithm
\ref{eq:two} the first arrow corresponds to the inversion of
formulas \ref{eq:rt} and \ref{eq:standard}. However, $q$ and $Z_B$
were already found in \cite{OBrien0,OBrien2} and one can use the
data presented in P. 56 of \cite{OBrien2} for $n=q/k_0$ and $Z_B$.
In the second step of \ref{eq:two} one can use formulas
\ref{eq:general} and \ref{eq:ZB} solving them as an equation for $G$
and $X$ and finally find local MP from \ref{eq:gamma}--\ref{eq:ZZZ}.

To calculate the MP in the third example the geometry of infrared
SRRs are not important since the input data are values of $R$ and
$T$. The period $a$ in this case is equal $a=600$ nm and the time
dependence is $\exp(-i\omega t)$, as it is adopted in optics. The
input data taken from \cite{OBrien2} are shown in Fig. \ref{fignew}
(a,b) and the results are presented in Fig. \ref{fignew} (c-f). The
difference between the results obtained for local and non-local
material parameters confirms the theory developed in the present
paper. Parameters extracted by a direct inversion of Fresnel
formulas \ref{eq:rt} and \ref{eq:standard} are non-local and
parameters obtained using the suggested theory satisfy to the
locality conditions.

\section{Conclusion}

In this paper it is demonstrated that local material parameters of
lattices can be introduced not only at very low frequencies but also
in the region of moderately low frequencies where MTM operate. These
local MP complement a pair of non-local MP which were obtained
before in known works. Though at moderate frequencies local MP allow
to solve boundary problems for MTM slabs only at expense of
introducing two transition layers at the sides of the slab, these MP
are very important. Only these MP do not depend on the incidence
angle and only they can be used to study the interaction of MTM with
wave packages and evanescent waves. It is then important to learn to
extract these parameters from $R$ and $T$ coefficients of MTM slabs.
The simplest algorithm of this extraction is explained in the
present paper, and the numerical examples show that the obtained MP
satisfy to locality conditions.

\end{document}